# Topology optimization of 2D structures with nonlinearities using deep learning

**Diab W. Abueidda** [a, b, c]**, Seid Koric** [a, b]**, Nahil A. Sobh** [a, c, *]

[a] National Center for Supercomputing Applications, University of Illinois at Urbana-Champaign

[b] Department of Mechanical Science and Engineering, University of Illinois at Urbana-Champaign

[c] Beckman Institute of Science and Technology, University of Illinois at Urbana-Champaign

**Abstract**

The field of optimal design of linear elastic structures has seen many exciting successes that resulted in new architected materials and structural designs. With the availability of cloud computing, including high-performance computing, machine learning, and simulation, searching for optimal nonlinear structures is now within reach. In this study, we develop convolutional neural network models to predict optimized designs for a given set of boundary conditions, loads, and optimization constraints. We have considered the case of materials with a linear elastic response with and without stress constraint. Also, we have considered the case of materials with a hyperelastic response, where material and geometric nonlinearities are involved. For the nonlinear elastic case, the neo-Hookean model is utilized. For this purpose, we generate datasets composed of the optimized designs paired with the corresponding boundary conditions, loads, and constraints, using a topology optimization framework to train and validate the neural network models. The developed models are capable of accurately predicting the optimized designs without requiring an iterative scheme and with negligible inference computational time. The suggested pipeline can be generalized to other nonlinear mechanics scenarios and design domains.

---

[*] Corresponding author: Nahil Sobh
 *E-mail address*: sobh@illinois.edu

1. Introduction

The pursuit of structures and materials with enhanced performance yet lightweight has been of high scientific and industrial interest [1-3]. Generally, such materials and structures can be obtained by selecting the constituents a) materials, b) volume fractions, and c) architectures. The former two approaches have been studied extensively and are almost mature [4]. On the other hand, designing the architectures of materials is still an active area of research, as it allows for obtaining unique properties [5-8]. The increased interest in architectured materials is related to their enhanced properties such as permeability, thermal and electrical conductivities, electromagnetic shielding effectiveness, stiffness-to-weight ratio, etc. [9, 10]. Recent advances in additive manufacturing have permitted the fabrication of such materials and structures with complex geometries [11-14]. Attaining architectures resulting in structures and materials with enhanced performance is usually based on intuitions, experiments, and/or bioinspiration [15, 16].

Topology optimization offers a systematic platform for obtaining new designs of materials and structural systems with optimized responses [17-23]. Generally, solving the inverse problem is a difficult task to deal with, in which specific parameters need to be found to obtain an optimal response, and to do so, the forward problem has to be solved iteratively [24], regardless of using gradient-based or gradient-free optimization algorithms. In topology optimization problems, one aims at identifying the optimal material distribution yielding the desired properties such as maximization of energy absorption and minimization of compliance, while still, the design constraints are satisfied. James et al. [25] developed a framework for optimizing structures where they accounted for material damage. The failure is mitigated by enforcing a constraint on the maximum local damage intensity. Also, Russ et al. [26] used the phase-field method for the fracture to increase the structural fracture resistance and strength. Geometrically nonlinear structures have also been studied, as shown in [27, 28].

Another intriguing problem in the field of topology optimization is problems involving many load cases. Zhang et al. [29] proposed a computationally-efficient randomized approach for deterministic topology optimization with many load cases. Lately, manufacturing-oriented topology optimization has experienced an increasing interest by both industry and academia, especially with recent advances in the field of additive manufacturing [30]. Also, increasing attention is observed for developing topology optimization algorithms for multi-material structures. For example, Alberdi et al. [31] developed a bi-material topology optimization framework, where hyperelastic and viscoplastic phases are combined, for maximizing energy dissipation. Additionally, Conlan-Smith et al. [32] applied topology optimization to design compliant mechanisms using functionally graded materials.

Generally, topology optimization problems are very computationally expensive due to a large number of design variables and the need for many optimization iterations before obtaining the optimal one [33]. Also, gradient-based topology optimization algorithms may suffer from the dependency on the starting point, given that multiple local optima exist. In such a scenario, it is probable that the attained optimized solution is not the global optimum. These drawbacks urge many researchers to develop more efficient frameworks to determine the optimal solution. For instance, Lee et al. [34] proposed a new meta-heuristic optimization algorithm suitable for engineering applications.

Advances in high-performance computer (HPC) hardware and scalable solver algorithms have revolutionized various science and engineering fields in the last two decades allowing high fidelity nonlinear finite element (FE) simulations of highly heterogeneous materials [35] as well as multiphysics even on the petascale computing architecture [36, 37]. The field of machine learning (ML) is no exception, and particularly deep learning has benefited from these technological advances, especially on graphics processing units (GPU). ML has been successful and effective in spam detection, image and speech recognition, discoveries of diseases and drugs, remote sensing image analysis for traffic applications, and search engines [38-40].

Furthermore, ML has shown success in mechanics-related fields [41-46], including and limited to predicting solidification defects [47] and effective thermal conductivities of composites [48, 49], solving multiphysics problems [50], and designing new materials [51, 52]. Bessa et al. [53] showed that obtaining material models using ML is possible, providing that the computational analyses of representative volume elements (RVEs) have high fidelity and enough efficiency required to generate sufficient data for supervised learning tasks. The use of ML algorithms has intriguingly been extended to the prediction and optimization of different materials and structural systems [54-63]. Also, neural networks have been used to solve partial differential equations (forward boundary value problems), avoiding the conventional discretization involved in the finite element method, by using either energy approach [64] or collocation strategy [65].

Abueidda et al. [66] developed a convolutional neural network (CNN) model that is capable of quantitatively predicting the stiffness, strength, and toughness of a two-dimensional (2D) checkerboard composite. Also, they integrated the CNN model with a genetic algorithm to solve

single- and multiple-objective optimization problems. The use of deep learning was taken one step further to precisely predict plasticity-constitutive laws as detailed in [67], in which the authors showed that sequence learning can obtain the evolution of stresses and plastic energy, given a deformation path.

Recently, deep learning has been implemented to perform optimization procedures directly without the need to involve an optimizer as in the work of Abueidda et al. [66] and Sasaki et al. [68]. This is accomplished by training the deep learning algorithms to produce images of the optimized designs given a set of boundary conditions and loads [69, 70]. For instance, Yu et al. [71] proposed a deep learning model that is capable of identifying optimal designs without using an iterative scheme. The model was trained on synthetic data generated by an open-source code for linear elastic optimization. Moreover, Rawad and Shen [72, 73] employed a generative adversarial network, which consists of a discriminator and a generator, to optimize two-dimensional (2D) and three-dimensional (3D) linear elastic structures. Also, Zhang et al. [74] developed a CNN model, composed of an encoder and decoder, that identifies the optimal designs in negligible time. The material they considered is a linear elastic one assuming infinitesimal strain theory. White et al. [75] developed a multiscale topology optimization framework for elastic structures using a neural network surrogate model.

So far, the implementation of machine learning algorithms in topology optimization has been limited to design spaces with linear elastic materials undergoing small deformation, with linear optimization constraints. Several studies have shown that geometric and material nonlinearities significantly influence the solution of the optimization, provided that the loads are large enough to onset system nonlinearities [76-78]. In this paper, we develop three CNN models to predict the material distribution possessing the optimized response, where the first model assumes linear elastic material and small deformations without stress constraint, while the second model accounts for large deformations. The CNN model accounting for large deformations is developed for materials obeying the hyperelastic neo-Hookean constitutive model. The third CNN model assumes a linear elastic material under a stress constraint [22, 79, 80]. The stress constraint is efficiently imposed using a smooth maximum function using global aggregation.

In this paper, we develop ML models that perform a real-time topology optimization of materials under large deformation and small deformation (with and without stress constraint). The remainder of the paper is organized as follows: Section 2 provides an overview of the general topology optimization problem we are interested in. Section 3 scrutinizes the sample space and associated training and testing datasets. Section 4 discusses the architectures of the CNN models and their corresponding hyperparameters and states the loss function and metrics employed in evaluating the performance of the CNN models. In Section 5, we present the results along with analysis and discussion. We conclude this study in Section 6 by summarizing the significant outcomes and discussing potential directions for future work.

## 2. Topology optimization

### 2.1. Linear and nonlinear structures

Generally, topology optimization algorithms attempt to identify the optimal material distribution within a given design space that minimizes or maximizes single or multiple objective function(s) while a set of constraints are satisfied. Topology optimization problems are solved by directly optimizing the location of the material boundary inside a design space [81], or they are solved by determining elements to be contained within a material region [25]. In this study, the latter approach is used along with the solid isotropic material penalization (SIMP) method [82]. In this study, the penalization factor is set to 3. Having the penalization factor larger than 1 penalizes the intermediate densities, so the algorithm converges to a solution of binary (0-1) densities. Following this approach, each finite element (in a finite element idealization of a structure) has a density attribute $\rho_e \in [0,1]$, and each element density is considered as a design variable in the optimization problem.

In the SIMP method, penalization factor $n$ is used to steer the densities $\rho$ to a value of zero or one. The parametrization is achieved by writing the total elastic energy $SE$ as

$$SE = \sum_{e=1}^{n_{ele}} \rho_e^n \int_{\Omega_e} \psi d\Omega_e \qquad (1)$$

where $\psi$ is the strain energy density function, $n_{ele}$ denotes the total number of elements, and $\Omega_e$ is the reference configuration of element $e$ [83]. The element-based formulation can suffer from numerical instabilities such as checkerboarding and mesh-dependence [84, 85], where density filtering technique [86] can be employed to address these issues. Here, we consider two types of strain energy density functions: 1) a linear elastic strain energy density function $\psi^{LE}$ and 2) a hyperelastic strain energy density function based on neo-Hookean material $\psi^{NH}$, where materials are assumed to be isotropic. The linear elastic strain energy density function is written as

$$\psi^{LE} = \frac{1}{2} C_{ijkl} \varepsilon_{ij} \varepsilon_{kl}$$
$$C_{ijkl} = \kappa \delta_{ij} \delta_{kl} + \mu \left( \delta_{ik} \delta_{jl} + \delta_{il} \delta_{jk} - \frac{2}{3} \delta_{ij} \delta_{kl} \right) \qquad (2)$$
$$\varepsilon_{ij} = \frac{1}{2} \left( u_{i,j} + u_{j,i} \right)$$

where $\varepsilon_{ij}$ denotes a component of an infinitesimal strain tensor, $C_{ijkl}$ is a component of the fourth-order elasticity tensor, $\delta_{ij}$ is the Kronecker delta, $u_i$ denotes a displacement component, and $(\bullet)_{,j}$ is the gradient operator. The material parameters $\kappa$ and $\mu$ represent the bulk and shear moduli, respectively. On the other hand, the neo-Hookean strain density function [87] is expressed as

$$\psi^{NH} = C_{10}(\bar{I}_1 - 3) + \frac{1}{D_1}(J-1)^2$$
$$\bar{I}_1 = \bar{\lambda}_1 + \bar{\lambda}_2 + \bar{\lambda}_3 \tag{3}$$
$$J = \det(\mathbf{F})$$

where $C_{10}$ and $D_1$ are material parameters, $\mathbf{F}$ is the deformation gradient, $\bar{I}_1$ denotes the first deviatoric strain invariant, and $\bar{\lambda}_i$ are the deviatoric stretches defined as $\bar{\lambda}_i = J^{-1/3}\lambda_i$ where $\lambda_i$ are the principal stretches. In the case of small deformation, $C_{10}$ and $D_1$ reduce to $C_{10} = \mu/2$ and $D_1 = 2/\kappa$.

In this paper, the objective function $G$ is defined as the compliance, sum of all elemental strain energies. G is minimized over a domain composed of a structure that is subject to prescribed boundary and loading conditions as well as volume constraint $V_f$. Mathematically, this optimization problem [86, 88] can be expressed as

$$\min_{\boldsymbol{\rho}} G(\boldsymbol{\rho}) = \mathbf{P}^T \mathbf{U}^f + \mathbf{R}^T \mathbf{U}^p,$$
$$subject\ to\ g_1 = \frac{V(\boldsymbol{\rho})}{V_o} - V_f \leq 0 \tag{4}$$

where **P** denotes the applied load vector, **R** is the reaction force vector, $(\bullet)^T$ represents the transpose operator, and $\mathbf{U}^f$ and $\mathbf{U}^p$ denote the unknown free and known prescribed displacement vectors, respectively. Also, $V$ is the volume of the design structure, and $V_o$ denotes the volume of the design space. Here, we use the optimization software package TOSCA [88, 89] callable from a general-purpose implicit finite element analysis (FEA) code [87] to perform the topology optimization tasks at hand. TOSCA has two algorithms for solving the optimization problems; these are known as the sensitivity-based solver and the controller-based solver. Both have their pros and cons, and the type of optimization problem at hand decides which one to use. We have used the sensitivity-based solver, which uses the method of moving asymptotes (MMA) [90, 91].

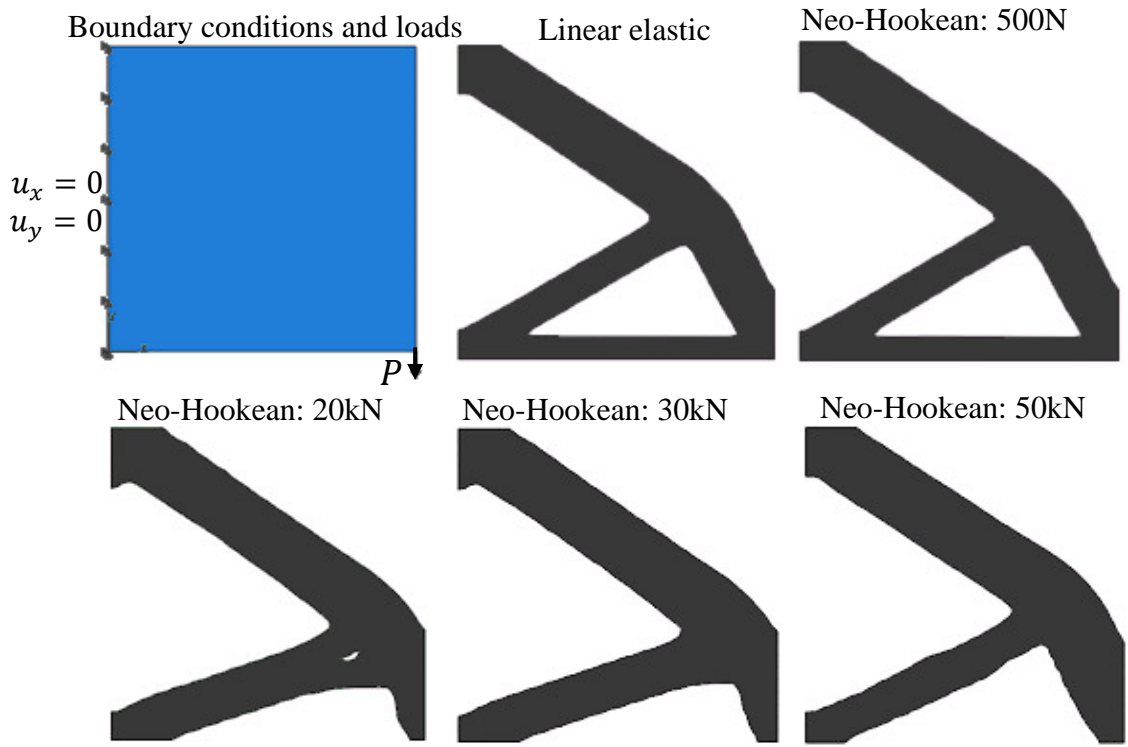

Figure 1: Illustration of boundary conditions and impact of load amplitudes on the final optimized design. Optimal designs have a volume constraint of 0.35.

As mentioned earlier, material and geometric nonlinearities impact the optimal design especially when applied loads are sufficiently large to trigger structural and/or materials nonlinearities [76-78]. The influence of nonlinearities, on the final optimal design, is exemplified below. Consider a 2D design space consisting of 2500 (50×50) elements, where the problem stated in equation (4) is solved with a volume constraint $V_f = 0.35$. The load and boundary conditions are as shown in Figure 1. The dimensions of the design space are $1m \times 1m$. In the case of large deformation and neo-Hookean model, the material parameters used are $C_{10} = 1 \text{MPa}$ and $D_1 = 1 \times 10^{-8} \text{Pa}^{-1}$ (such material constants are representative of rubber mechanical properties [92]), while in the case of small deformations and linear elasticity, the materials parameters are $\mu = 2C_{10} = 2 \text{MPa}$ and $\kappa = 2/D_1 = 200 \text{MPa}$. Figure 1 depicts the optimal designs for the elastic material with small deformations and neo-Hookean material with large deformations. The optimal design for linear elastic material with small deformation is independent of load amplitude, provided the direction is fixed. At small loads, the optimized designs obtained using the neo-Hookean hyperelasticity are identical/similar to the one attained from the linear elastic structures undergoing small deformations. However, changes in the optimized design take place when the load magnitude is increased when the neo-Hookean model is considered.

### *2.2. Nonlinear stress constraint*

In addition to the optimization scenarios discussed above, we also consider the case of linear elastic material under stress constraint [22, 79, 80], a nonlinear optimization constraint. Here, a single q-norm aggregation to the element average von Mises stress, over the entire design space, is applied:

$$\min_{\boldsymbol{\rho}} G(\boldsymbol{\rho}) = \mathbf{P}^T \mathbf{U}^f + \mathbf{R}^T \mathbf{U}^p,$$

$$\text{subject to } g_1 = \frac{V(\boldsymbol{\rho})}{V_o} - V_f \leq 0 \tag{5}$$

$$g_2 = \left[ \sum_{e=1}^{n_{ele}} \left( \rho_e^\eta \frac{\bar{\sigma}_{VM,e}}{\sigma_{lim}} \right)^q \right]^{1/q} - (1+\xi) \leq 0,$$

where $\bar{\sigma}_{VM,e}$ is the element average von Mises stress, $\sigma_{lim}$ is the allowable stress limit, and $\xi$ is a small number $(1 \times 10^{-7})$ used for numerical purposes. Such a constraint plays a vital role in efficiently mitigating structural failure by aggregating the stress constraints into one global constraint rather than locally enforcing stress constraints (element-level). The aggregation exponent $q$ is assigned a value of $q = 10$, while the stress relaxation exponent $\eta$ is selected to be $\eta = 3$. For a 2D problem, $\sigma_{VM}$ is defined as:

$$\sigma_{VM}^2 = \sigma_x^2 + \sigma_y^2 + \sigma_x \sigma_y + 3\sigma_{xy} \tag{6}$$

where $\sigma_x$, $\sigma_y$, and $\sigma_{xy}$ are the different stress components. The FE formulation for the element average von Mises stress is expressed as

$$\bar{\sigma}_{VM,e}^2 = \frac{1}{A_e} \boldsymbol{u}_e^T \left( \sum_{Gauss} \boldsymbol{J} \boldsymbol{B}^T \boldsymbol{E} \boldsymbol{M} \boldsymbol{E} \boldsymbol{B} \right) \boldsymbol{u}_e \tag{7}$$

where $A_e$ is the element area, $\boldsymbol{u}_e$ is the element displacement, $\boldsymbol{J}$ is the Jacobian, and $\sum_{Gauss}$ represents numerical integration by Gauss quadrature. $\boldsymbol{E}$ is the plane stress material matrix, $\boldsymbol{B}$ is the traditional shape function derivative matrix, and $\boldsymbol{M}$ is the coefficient matrix (inferred from equation (6)) defined as

$$M = \begin{bmatrix} 1 & -\frac{1}{2} & 0 \\ -\frac{1}{2} & 1 & 0 \\ 0 & 0 & 3 \end{bmatrix}. \tag{8}$$

Assuming uniform mesh and applying the SIMP method ($n$ is the penalization factor), $\bar{\sigma}_{VM,e}$ is written as

$$\bar{\sigma}_{VM,e} = \rho_e^n \Delta_e^{1/2}, \quad \Delta_e = u_e^T k_{VMo} u_e$$
$$k_{VMo} = \frac{1}{A_e} \left( \sum_{Gauss} JB^T E_o M E_o B \right) \tag{9}$$

where $E_o$ is the material matrix with $\rho_e = 1$. Hence, the stress constraint shown in equation (5) is rewritten as

$$g_2 = Z^{\frac{1}{q}} - (1+\xi) \leq 0$$
$$Z = \sum_{e=1}^{n_{ele}} \left( \rho_e^{\eta+n} \frac{\Delta_e^{1/2}}{\sigma_{lim}} \right)^q \tag{10}$$

Then, we find the sensitivities of the stress constraint using the adjoint method,

$$\Pi = g_2 + \omega^{p^T} R^p + \omega^{f^T} R^f$$
$$R^p = K^{pf} u^f + K^{pp} u^p - P^p \tag{11}$$
$$R^f = K^{ff} u^f + K^{fp} u^p - P^f$$

where $K^{ij}$ is the partitioned (based on free and prescribed degrees of freedom) blocks of the global stiffness matrix, and $\omega^p$ and $\omega^f$ are the adjoint vectors associated with the prescribed and free degrees of freedom, respectively. Taking the implicit derivative $\Pi$ with respect to $\rho$ and performing some algebraic manipulation yield

$$\frac{D\Pi}{D\rho} = \frac{\partial g_2}{\partial \rho} + \omega^{p^T} \frac{\partial \mathbf{R}^P}{\partial \rho} + \omega^{f^T} \frac{\partial \mathbf{R}^f}{\partial \rho}$$
$$+ \left( \frac{\partial g_2}{\partial \mathbf{u}^f} + \omega^{p^T} \frac{\partial \mathbf{R}^P}{\partial \mathbf{u}^f} + \omega^{f^T} \frac{\partial \mathbf{R}^f}{\partial \mathbf{u}^f} \right) \frac{\partial \mathbf{u}^f}{\partial \rho} \quad (12)$$
$$+ \left( \frac{\partial g_2}{\partial \mathbf{P}^p} + \omega^{p^T} \frac{\partial \mathbf{R}^P}{\partial \mathbf{P}^p} + \omega^{f^T} \frac{\partial \mathbf{R}^f}{\partial \mathbf{P}^p} \right) \frac{\partial \mathbf{P}^p}{\partial \rho} \; .$$

Now, we select $\omega^p$ and $\omega^f$ such that the brackets in the second and third terms of equation (12) are zeros:

$$\frac{\partial g_2}{\partial \mathbf{u}^f} + \omega^{p^T} \frac{\partial \mathbf{R}^P}{\partial \mathbf{u}^f} + \omega^{f^T} \frac{\partial \mathbf{R}^f}{\partial \mathbf{u}^f} = \mathbf{0}$$
$$\frac{\partial g_2}{\partial \mathbf{P}^p} + \omega^{p^T} \frac{\partial \mathbf{R}^P}{\partial \mathbf{P}^p} + \omega^{f^T} \frac{\partial \mathbf{R}^f}{\partial \mathbf{P}^p} = \mathbf{0} \quad (13)$$

From equations (11) and (13), $\omega^p = \mathbf{0}$. Combining equations (11), (12), and (13) yields

$$\frac{D\Pi}{D\rho} = \frac{\partial g_2}{\partial \rho} + \omega^T \frac{\partial \mathbf{R}}{\partial \rho}$$
$$\omega^p = \mathbf{0}, \quad \mathbf{K}^{ff} \omega^f = -\frac{\partial g_2}{\partial \mathbf{u}^f}. \quad (14)$$

From equation (14), one needs to find the explicit derivatives $\frac{\partial g_2}{\partial \mathbf{u}^f}$ and $\frac{\partial g_2}{\partial \rho}$ to obtain the implicit derivative $\frac{D\Pi}{D\rho} = \frac{Dg_2}{D\rho}$:

$$\frac{\partial g_2}{\partial u_i} = Z^{\frac{1-q}{q}} \sum_{e=1}^{n_{ele}} \left[ \left( \frac{\rho_e^{\eta+n}}{\sigma_{lim}} \right)^q \Delta_e^{\frac{q}{2}-1} (k_{VMo})_k \, u_e \right]$$

$$\frac{\partial g_2}{\partial \rho_e} = (\eta+n) \rho_e^{[(\eta+n)q-1]} Z^{\frac{1-q}{q}} \left( \frac{\Delta_e^{1/2}}{\sigma_{lim}} \right)^q$$

(15)

where $e$ is the element index, $i$ is the degree of freedom index, and $(k_{VMo})_k$ is the $k^{th}$ row in the matrix $k_{VMo}$, where $k$ is the position in the element displacement $u_e$ corresponding to $u_i$.

We do not show the derivations of the objective function and volume constraint sensitivities, as they are popular in the topology optimization fields. Here, we solve the topology optimization using an in-house MATLAB code, in which the method of moving asymptotes [90, 91] is used to solve the optimization problem. The method of moving asymptotes creates analytical convex approximations of the nonlinear functions. It is crucial to verify that the sensitivities are correctly calculated before starting the generation of data. To do so, we compare the sensitivities obtained from the adjoint method with those obtained using the finite difference method, where small design space is considered for the verification purpose. The same boundary conditions and load location and angle as those shown in Figure 1 are used. The design space $(1m \times 1m)$ is discretized into 400 $(20 \times 20)$ elements. Please note that this mesh size is used only for the process of verifying the analytical sensitivities, and a finer mesh is used for generating the data. The material considered for the structure is epoxy: Young's modulus is $4.07$ GPa, Poisson's ratio is $0.34$, and allowable stress limit (yield stress) is $16.44$ MPa. In other words, we try to obtain an optimal design, given a set of loading conditions, without experiencing any plastic deformation. The filter radius considered is $0.1m$, and the force magnitude is $1$ MN $(M = 1 \text{ million})$. Choosing small maximum force values will naturally eliminate the effect of the nonlinear stress constraint, as such a value will not induce high-stress levels. A uniform element density of $\rho_e = 0.35$ is used for the sensitivity verification study. Figure 2 shows the comparison between the sensitivities obtained using the adjoint and finite difference methods. Both constraints are non-dimensional, while the objective function has a unit of Joule.

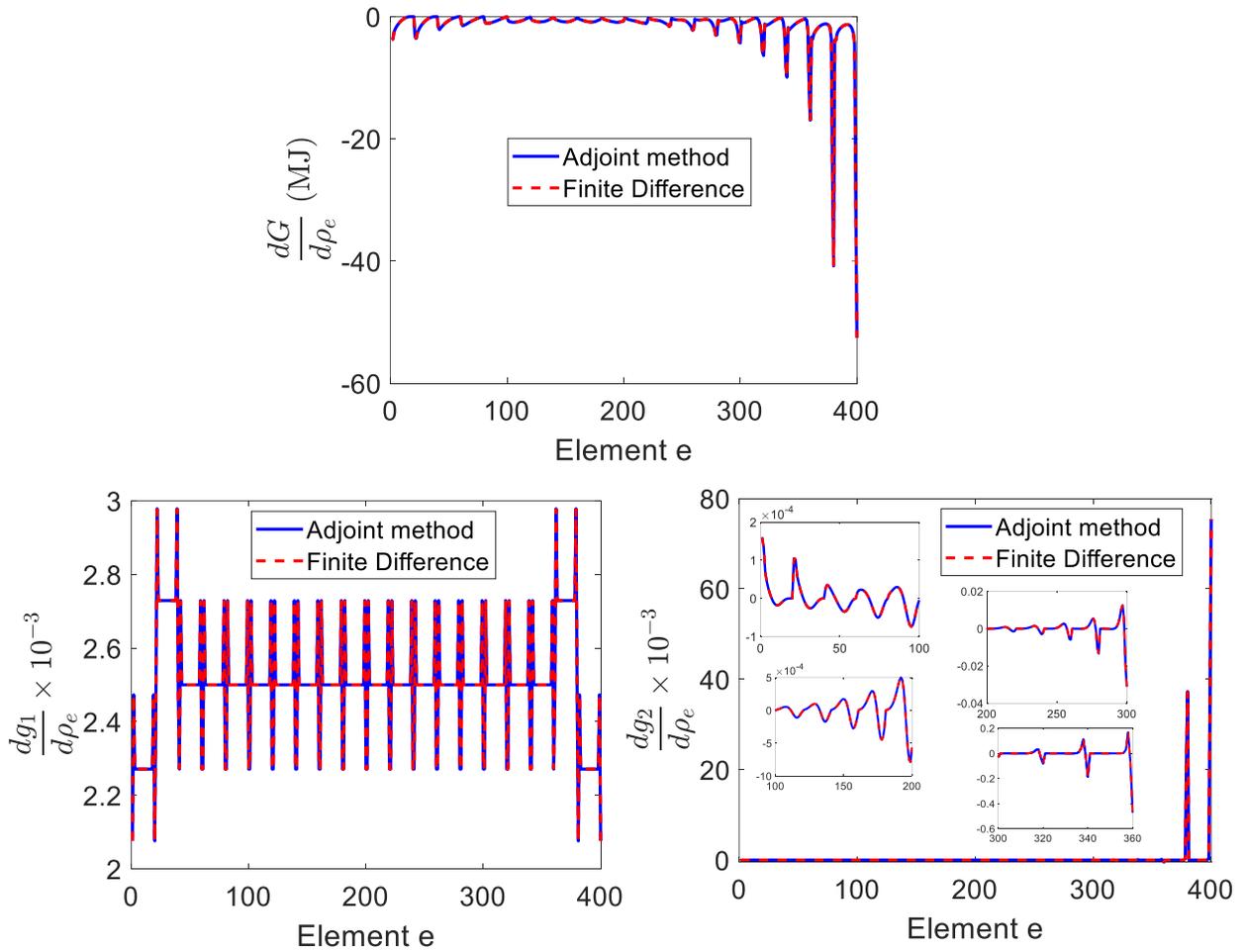

Figure 2: Verification of sensitivity analyses.

3. **Data description, generation, and processing**

   *3.1. Elasticity and hyperelasticity*

Here, two CNN models are developed, one for a linear elastic material and another for a neo-Hookean rubber-like material. A dataset is generated for each of these two material models. Each dataset is composed of many pairs of optimized designs and their corresponding boundary conditions, loads, and volume constraints. In this study, the proposed framework is illustrated using a single concentrated force (at a node on the right-hand side of the design space) while fixed displacements are imposed (at all nodes located at the left-hand side of the design space). The material properties used to generate the datasets are the same as the ones mentioned in Section 2.1.

For the sake of comparison with the work of Yu et al. [71], a 32 × 32 finite element mesh is used to discretize the linear elastic structure. The position, angle of incidence and the volume fraction are from a uniform random distribution with the following ranges: 1) the location of the applied force, the node selected from the set of the nodes at the right-hand side of the design space 2) the angle of the applied force ranges $\left(\theta \in [0, 2\pi]\right)$, and 3) the volume constraint ranges $\left(V_f \in [0.2, 0.8]\right)$. The filter radius $r_{min}$ is assigned a constant value of $r_{min} = 12.5\, cm$. For each data point in the dataset, the three parameters are randomly selected using uniform distribution functions available in the open-source package Python. Then, these parameters are automatically fed to the ABAQUS environment to generate the mesh, assign the boundary conditions and loads, define material properties, create the optimization problem with desired optimization parameters, and find the corresponding optimized design. Afterward, the optimized designs are saved to a text file, including all required information (input and output).

A total of 15,000 data pairs were generated using the iForge HPC cluster hosted at the National Center for Supercomputing Applications (NCSA). iForge consists of Intel/Skylake nodes, each with 40 cores and 192 GB of RAM, and a couple of nodes are also equipped with NVIDIA

v100 GPU cards. High throughput computing is applied to generate as many as ten data points simultaneously with the average rate of data generation of 0.31 minutes/data point. On a personal computer with CORE i5 vPro, a single optimization task takes roughly 25 minutes to be generated. The size of the dataset is determined by the performance of the model. The convergence condition for settling the size of the generated dataset is to achieve a dice similarity coefficient $(DSC)$ higher than 0.95. The interpretation of the $DSC$ is discussed in more detail in Section 4.2.

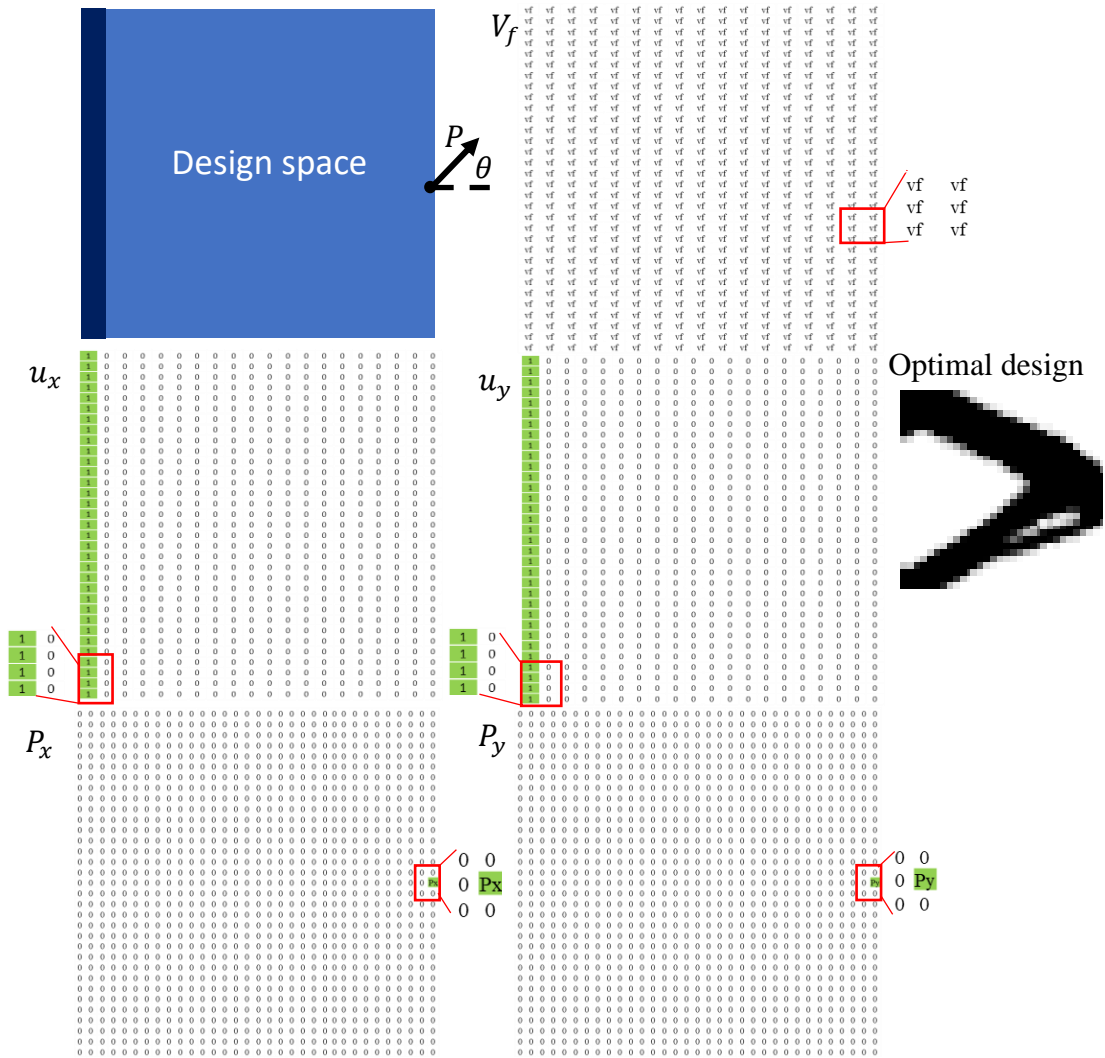

Figure 3: Demonstration of the different channels.

The next step is to arrange the generated data into a form suitable for the CNN model. With the current selection of design space and number of elements ($32 \times 32$), we have $33 \times 33$ nodes. Each the input of each data point can be viewed as five channels (images): 1) $u_x$ with a dimension of $33 \times 33$, 2) $u_y$ with a dimension of $33 \times 33$, 3) $P_x$ with a dimension of $33 \times 33$, 4) $P_y$ with a dimension of $33 \times 33$, and 5) $V_f$ with a dimension of $32 \times 32$. $u_x$ and $u_y$ matrices have zero components everywhere except at the nodes at the left-hand side, where fixed boundary conditions are imposed, a value of 1 is assigned. $P_x$ and $P_y$ matrices have zero values everywhere except at the node having the load $P$ applied. As discussed in the previous section, the magnitude of the load, when linear elastic material with small deformation is considered, does not affect the optimized design. Hence, $P_x$ and $P_y$ are computed as $P_x = cos\theta$ and $P_y = sin\theta$. Regarding the fifth channel, we adopt a different approach to include the information about the desired volume constraint than the approach Yu et al. [71] had used. In our approach, we use a separate channel with a uniform value of $V_f$ as part of the input, while Yu et al. [71] are passing the volume constraint information to the latent variable as a scalar input. On the other hand, the output of each data is composed of one channel, where the values of the different pixels (elements) are the densities obtained from the optimization framework. Having said that, the pixels of all input and output channels have values ranging between zero and one. Figure 3 portrays an example of a data point; Figure 3 shows the different channels.

For the neo-Hookean model, the considered design space has a dimension of $1m \times 1m$, where the design space has been meshed with $50 \times 50$ elements. Four parameters have been varied: 1) the location of the applied force, which node at the right-hand side of the design space has the load applied, 2) the magnitude of the load applied $P \in [0, P_{max} = 150,000\,\text{N}]$, 3) the angle of the applied force $(\theta \in [0, 2\pi])$, and 4) the volume constraint $(V_f \in [0.2, 0.8])$. The filter radius $r_{min}$ is assigned a constant value of $r_{min} = 8\,cm$. Like the linear elastic case, the varied parameters are randomly selected using a uniform distribution. The convergence condition for determining the size of the dataset is to achieve a $DSC$ higher than 0.95. Eighteen thousand data points had been generated to train and test the developed CNN model. In the case of hyperelasticity and large deformation, the average rate for data generation is 3.2 minutes/data point, having ten optimization tasks being solved simultaneously. Solving a single optimization task on a personal computer with CORE i5 vPro takes around 90 minutes to be completed. Figure 4 shows the data generation rates for the linear and nonlinear problems.

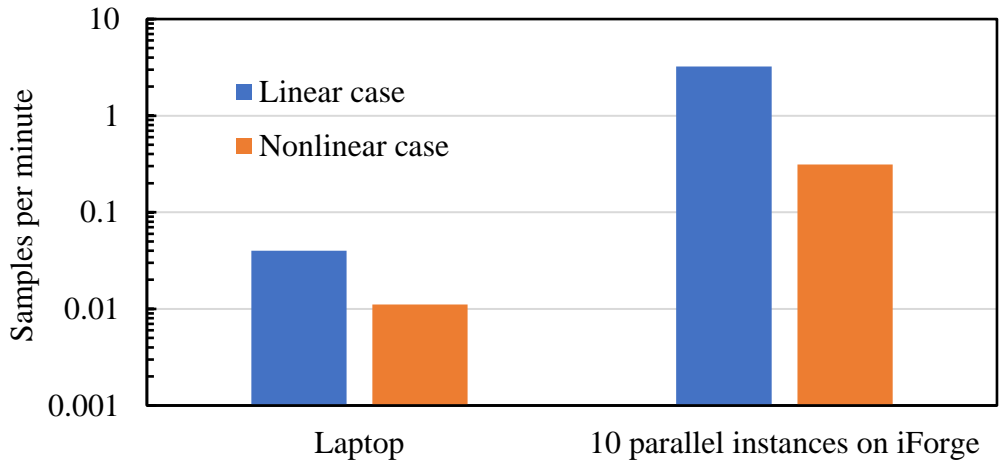

Figure 4: Data generation rates for the linear and nonlinear case.

The next step is to arrange the generated data into a form suitable for the CNN model. The procedure is very similar to the case of the linear elastic case. The design space has $50 \times 50$ elements and $51 \times 51$ nodes. The five channels are: 1) $u_x$ with a dimension of $51 \times 51$, 2) $u_y$ with a dimension of $51 \times 51$, 3) $P_x$ with a dimension of $51 \times 51$, 4) $P_y$ with a dimension of $51 \times 51$, and 5) $V_f$ with a dimension of $50 \times 50$. $u_x$ and $u_y$ matrices are initialized with zero value, and then values of 1 are assigned at the nodes at the left-hand side, where fixed boundary conditions are imposed. $P_x$ and $P_y$ matrices are initialized with zero values, and then nonzero values are assigned at the node having the load $P$ applied. The values of the pixels corresponding to the node having the load $P$ applied are $P_x = P\frac{\cos\theta}{P_{max}}$ and $P_y = P\frac{\sin\theta}{P_{max}}$. A uniform value of $V_f$ is assigned for the volume fraction channel. The output of each data is composed of one channel, where the pixels have values equal to the densities obtained from the optimization framework.

### 3.2. Nonlinear stress constraint

The data used to train the stress-based topology optimization CNN model are generated using an in-house MATLAB code. The considered design space has a dimension of $1m \times 1m$, where the design space has been discretized into $50 \times 50$ elements. The base material for the structure is the same as the one discussed in Section 2.2. Unlike the case of linear elasticity without stress constraint, the magnitude of the force affects the optimized design due to the incorporation of the stress constraint. Also, we take into consideration the effect of the filter radius. Five parameters have been varied: 1) the location of the applied force, which node at the right-hand side of the design space has the load applied, 2) the magnitude of the load applied $P \in [0, P_{max} = 1\text{MN}]$, 3) the angle of the applied force $(\theta \in [0, 2\pi])$, 4) the volume constraint $(V_f \in [0.2, 0.8])$, and the filter radius $(r_{min} \in [3cm, 10cm])$. Like the previous two cases, the varied parameters are randomly selected using a uniform distribution. The convergence condition for determining the size of the dataset is to achieve a $DSC$ higher than 0.95. Twenty thousand data points had been generated to train and test the developed CNN model.

Then, we arrange the generated data into a form suitable for the CNN model. The design space has $50 \times 50$ elements and $51 \times 51$ nodes. The six channels are: 1) $u_x$ with a dimension of $51 \times 51$, 2) $u_y$ with a dimension of $51 \times 51$, 3) $P_x$ with a dimension of $51 \times 51$, 4) $P_y$ with a dimension of $51 \times 51$, 5) $V_f$ with a dimension of $50 \times 50$, and 6) $r_{min}$ with a dimension of $50 \times 50$. The first five channels are created using the same approach utilized in creating the input channels in the case of the neo-Hookean material discussed in Section 3.1. In addition to these five channels, we have an extra channel accounting for the filter radius, where all pixels in this channel are assigned a uniform value $r_{min}$. The output of each data is composed of one channel, where the pixels have values equal to the densities obtained from the optimization framework.

Although one can arrange the data (channels) for the three scenarios we have considered (linear elasticity with and without stress constraint and large-deformation hyperelasticity) in other ways, we stick with this approach as it makes clear how one can generalize the CNN model, so it accounts for scenarios where the prescribed displacements and forces can be on different edges. Also, the adopted CNN model [40] requires the inputs and outputs to have a size of $2^m \times 2^m$, where $m$ is a positive integer. Hence, padding is done, so all the channels (inputs and outputs) have a size of $64 \times 64$ pixels. For all cases (linear elasticity with and without stress constraint and hyperelasticity), the images can be cropped to remove the padding and retrieve the original size of each problem.

Figure 5: Illustration of the a) building block used in U-net, b) building used in ResUnet, and c) architecture of the ResUnet.

## 4. ResUnet

*4.1. ResUnet architecture*

The primary objective of this paper is to develop deep CNN models to solve topology optimization problems. The adopted CNN model is based on the ResUnet proposed by Zhang et al. [40]. The ResUnet is a semantic segmentation convolutional neural network combining the privileges of the U-net and residual learning to improve the performance of U-net further. U-net was initially proposed by Ronneberger et al. [93]. U-net concatenates feature maps from different levels to improve segmentation accuracy. In other words, U-net combines low-level detail information and high-level semantic information to enhance segmentation accuracy. This concatenation of feature maps from different levels is not utilized in the CNN model developed by Yu et al. [71].

Generally, deeper neural networks can help get models with better performance [94]. However, very deep neural networks encounter problems such as vanishing gradients. He et al. [95] presented a deep residual learning framework to facilitate the training of very deep networks. The primary difference between the employed ResUnet [40] and conventional U-net [93] is the use of residual units instead of plain neural units as building blocks for the developed network. Figure 5a and Figure 5b portray the building blocks used in the U-net and ResUnet, respectively. A residual unit is a combination of batch normalizations (BN), rectified linear units (ReLU), and convolutional layers (Conv).

Figure 5c depicts the architecture of the ResUnet. The ResUnet is composed of three components: 1) encoder, encodes input images into compact representation, 2) decoder, retrieves the encoded representations to a pixel-wise categorization (semantic segmentation), and 3) bridge, connects the encoder and decoder. The skip connections between the encoder and decoder and within the residual units ease information propagations in forward and backward directions and reduce the number of parameters needed. The reader is referred to the paper by Zhang et al. [40] for a more in-depth discussion about the network. It is worth highlighting that we have added one residual block to the encoder and its corresponding block to the decoder, as the original ResUnet architecture suggested in the paper is not deep enough to predict the optimized designs for the nonlinear case, and it is sufficient for the elastic case. To have a unified framework, we used the same number of residual blocks for the linear and nonlinear cases, although the linear case does not require any modification to the original architecture.

*4.2. Loss function and model evaluation*

We developed three ResUnet networks, one for the small-deformation linear elastic material with and without a nonlinear constraint and one for the neo-Hookean material with nonlinearities. The models were developed and tested using Keras [96]. Also, we utilize mini-batching to increase the convergence rate and assist the CNN models to escape from local minima [97]. The same hyperparameters are used for the three cases: the batch size of 64, the number of epochs of 150, and the learning rate of 0.001. We use Adam optimizer [98], which is a gradient-based stochastic optimization algorithm to train the models. The goal of the optimization problem is to find the weights $W$ of the network that minimize the loss between the ground-truth segmentation $s_i$ given input images $I_i$ and the segmentations generated by the network $Net(I_i;W)$. Here, we use the mean square error as our loss function

$$MSE = \frac{1}{N}\sum_{i=1}^{N}\left\|Net(I_i;W) - s_i\right\|^2 \qquad (16)$$

where $N$ is the number of training examples. Throughout the training process, another metric is monitored in addition to the history of the MSE loss. The dice similarity coefficient $(DSC)$ [99] is computed to evaluate the performance of the model and check its convergence. The $DSC$ measures the similarity between two images $[y, \bar{y}]$, where $y$ is the ground-truth image, and $\bar{y}$ is the predicted one. The $DSC$ used is expressed as

$$DSC = \frac{2|y \cap \bar{y}|}{|y| + |\bar{y}|}. \tag{17}$$

If two images are identical, the coefficient is equal to 1.0, while in the case of no common pixels between two images, the $DSC$ is equal to 0.0.

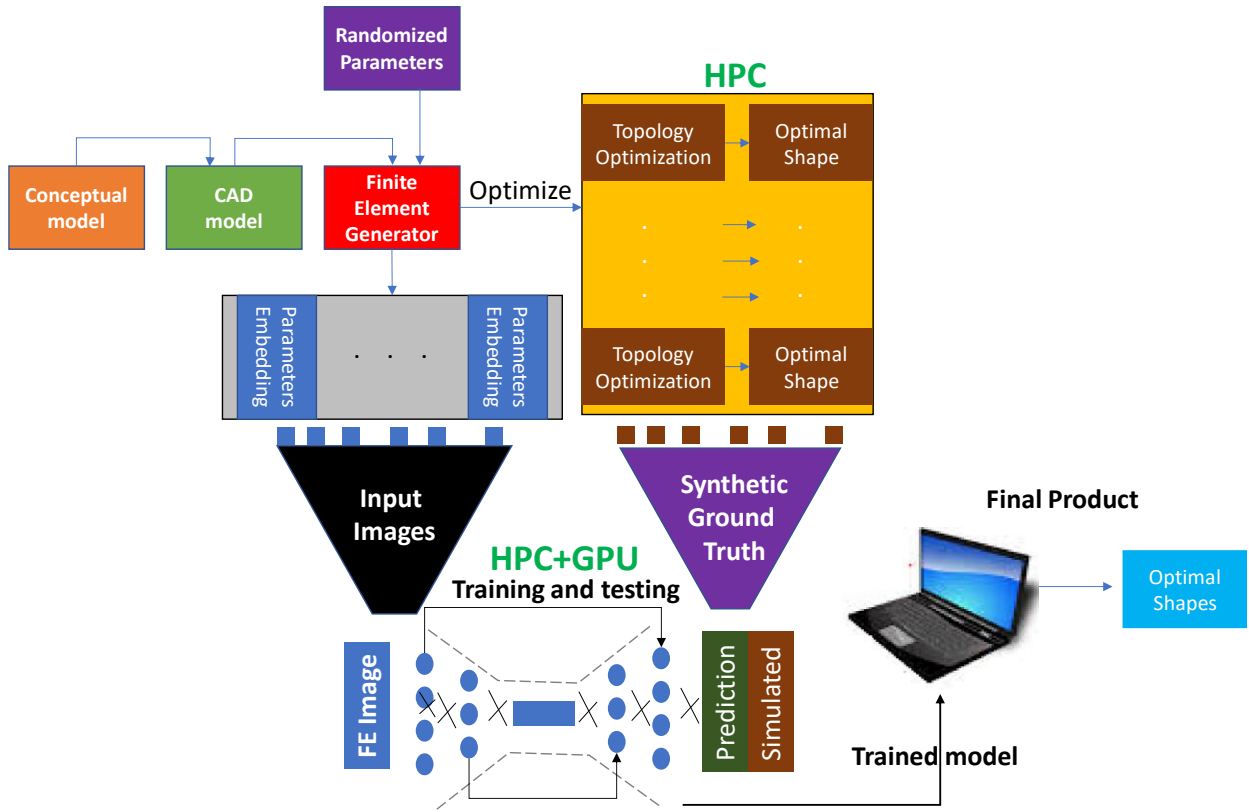

Figure 6: Flowchart showing the different steps used to develop a CNN-based optimizer.

## 5. Results and discussion

Figure 6 presents a flowchart showing the different stages of model development. The training of a CNN model is achieved by solving an optimization problem aiming at finding the parameters of the CNN model, so the loss function MSE is minimized. The CNN models developed for the linear elasticity (small deformation) with and without stress constraint and hyperelasticity are trained using 150 epochs. The data generated for each case are split into training (81%), validation (9%), and testing (10%) datasets. The training dataset is the dataset used to solve the optimization problem and find the parameters of the CNN model. The validation dataset is a set of data not used to find the optimized parameters of the CNN model, but they are used to evaluate the convergence progress of the model. After each epoch, the losses obtained from the validation and training datasets are compared. The testing dataset is used after the training process is completed to provide a final evaluation for the performance of the model.

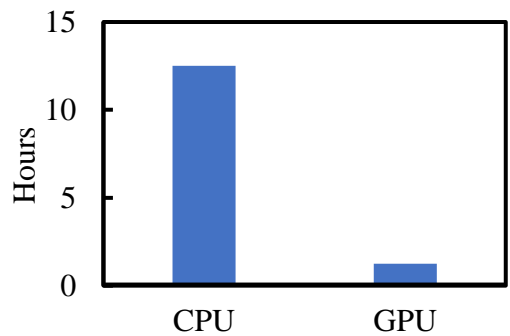

Figure 7: Demonstration of training time for the linear case when CPU-only iForge node with Skylake cores and a single GPU are used.

The training is done on a GPU node of iForge using a single NVIDIA v100 GPU card equipped with 32 GB of device memory. The training process takes 1.25 hrs and 1.5 hrs for the linear and nonlinear cases, respectively. Also, for comparison purposes, we run the linear case on a CPU-only iForge node with Skylake cores; the training requires 12.5 hrs, thus making an order of magnitude performance improvement on the GPU hardware. Figure 7 visualizes the training time required for the linear case when CPU-only and GPU nodes are used. Since the v100 GPU architecture has 4 GPU cards, a further performance acceleration is possible with the multi-GPU programming models, particularly with larger training data sizes, making GPU a much better choice in machine learning training.

Figure 8 demonstrates the convergence history of the loss function MSE for the cases of linear elasticity with small deformation and case of geometric and material nonlinearities. For both cases, the difference between the validation and training losses is small, and this indicates that overfitting is within an acceptable level. Also, the mean $DSC$ for the validation and training datasets are computed at the end of each epoch to evaluate the model. Figure 9 shows the convergence history of the $DSC$. After the completion of the training process, the testing dataset, which is different from the validation dataset and not seen by the model at all through the training dataset, is used to provide a final evaluation for the developed model. The evaluation is done using the testing dataset, and it is done quantitatively by computing the mean $DSC$ and qualitatively by randomly picking optimized designs from the testing dataset to compare between the ground-truth designs and predicted ones.

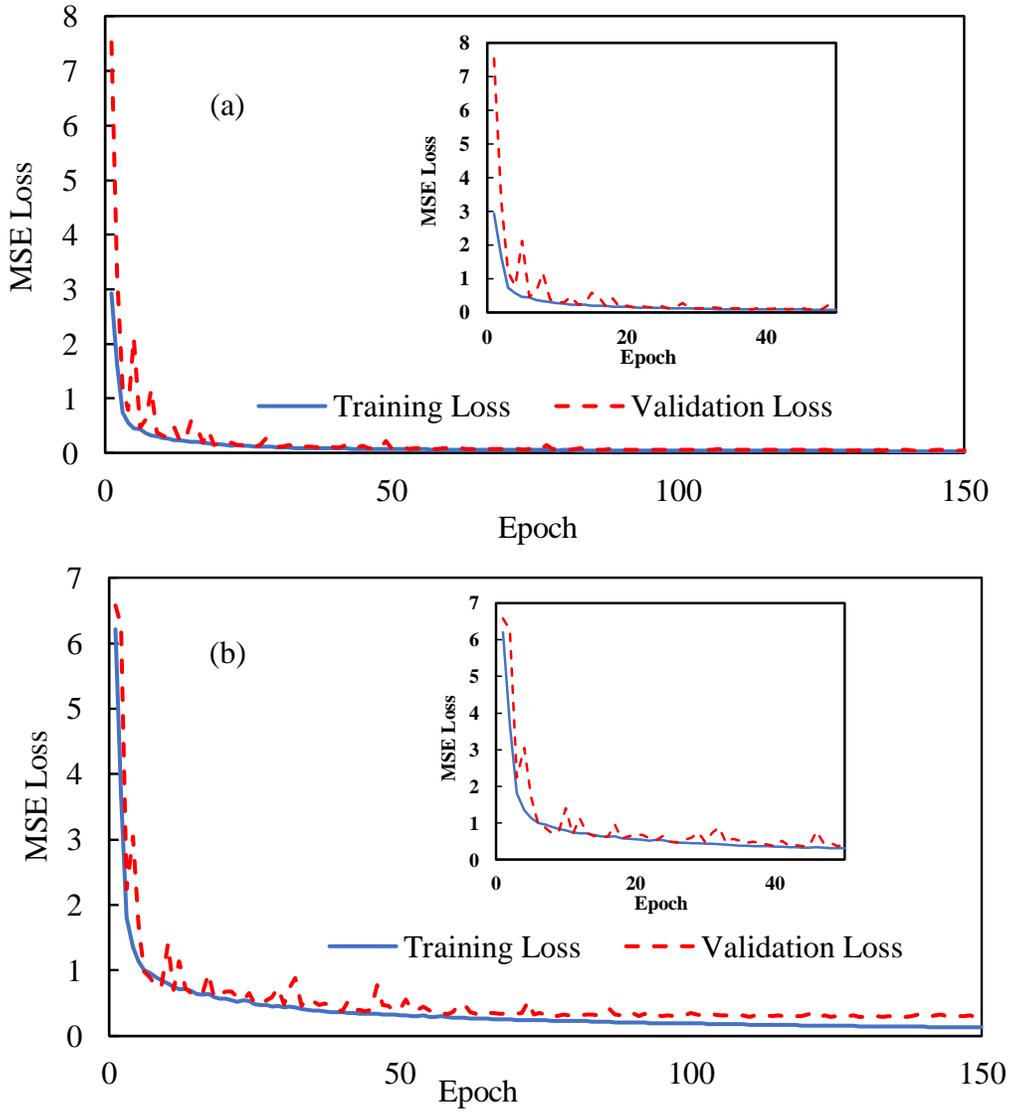

Figure 8: The convergence history of the loss function for the CNN model developed for the case of (a) linear elasticity with small deformation and (b) geometric and material nonlinearities.

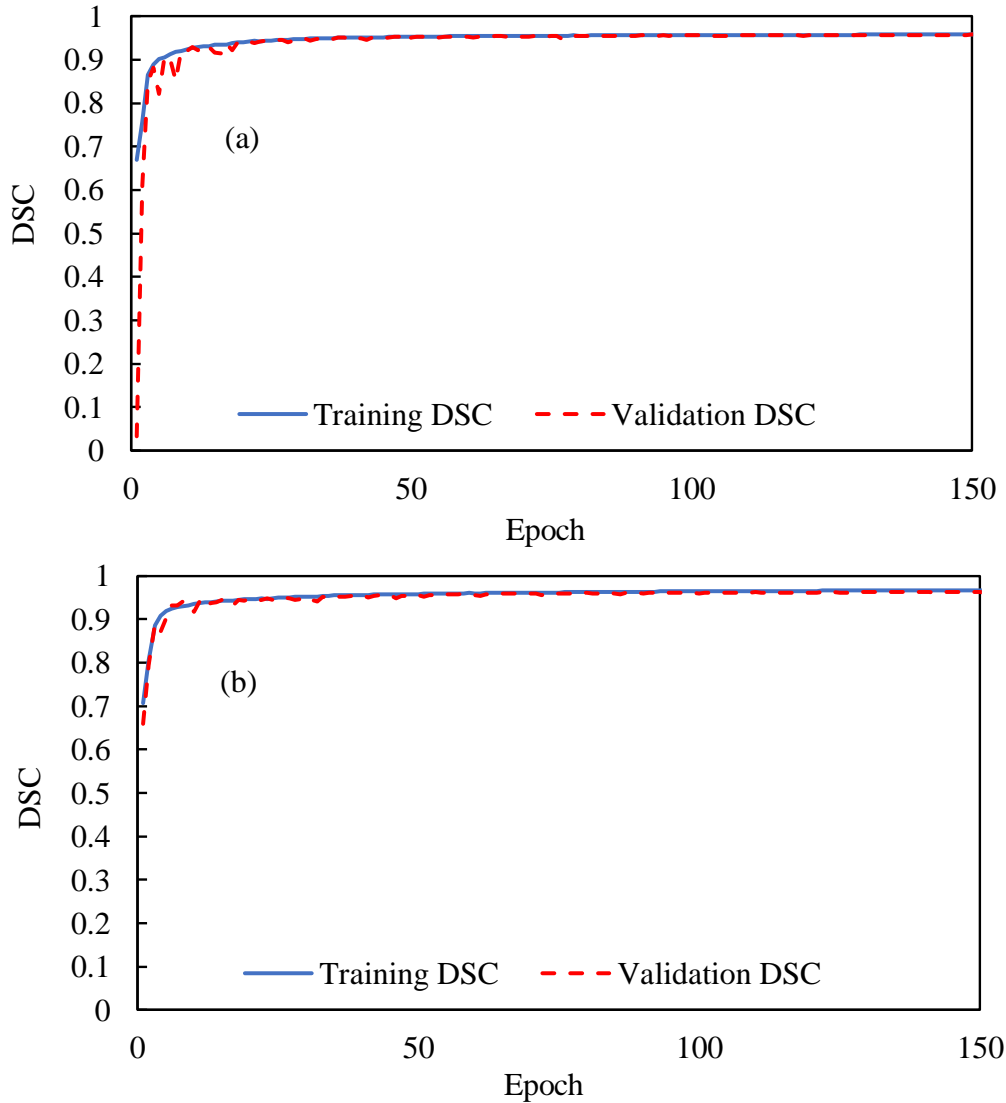

Figure 9: The convergence history of the $DSC$ for the CNN model developed for the case of (a) linear elasticity with small deformation and (b) geometric and material nonlinearities.

Conceptually, the resulted elemental densities range from 0 to 1. Here, a threshold value of 0.5 is used to retrieve the binary nature of the solution. After the training process is accomplished, densities with values larger than 0.5 are set to 1, while densities with values smaller than 0.5 are set to 0. Let's start with discussing the performance of the model developed for the linear elastic case (small deformation) without stress constraint. The mean $DSC$ is calculated using the testing dataset; the mean $DSC = 0.958$. This indicates that the ground-truth and predicted designs are almost identical, implying that the network is robust.

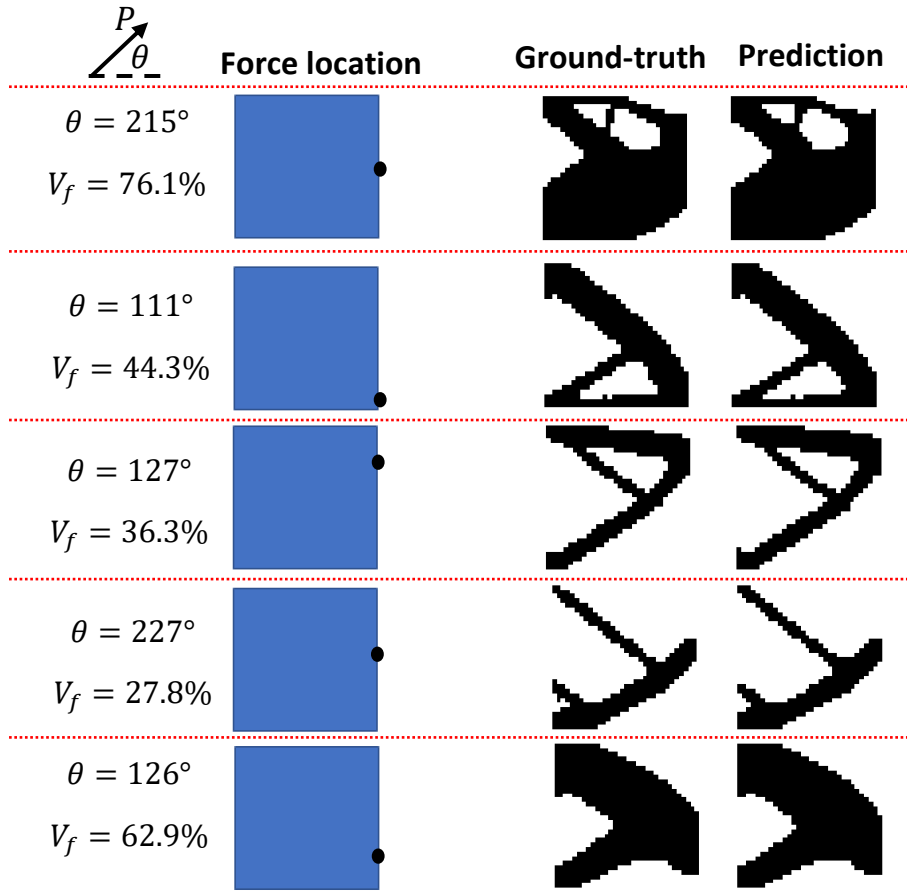

Figure 10: Comparison between optimized designs for the case of linear elasticity with small deformation. The design space has a dimension of $1m \times 1m$.

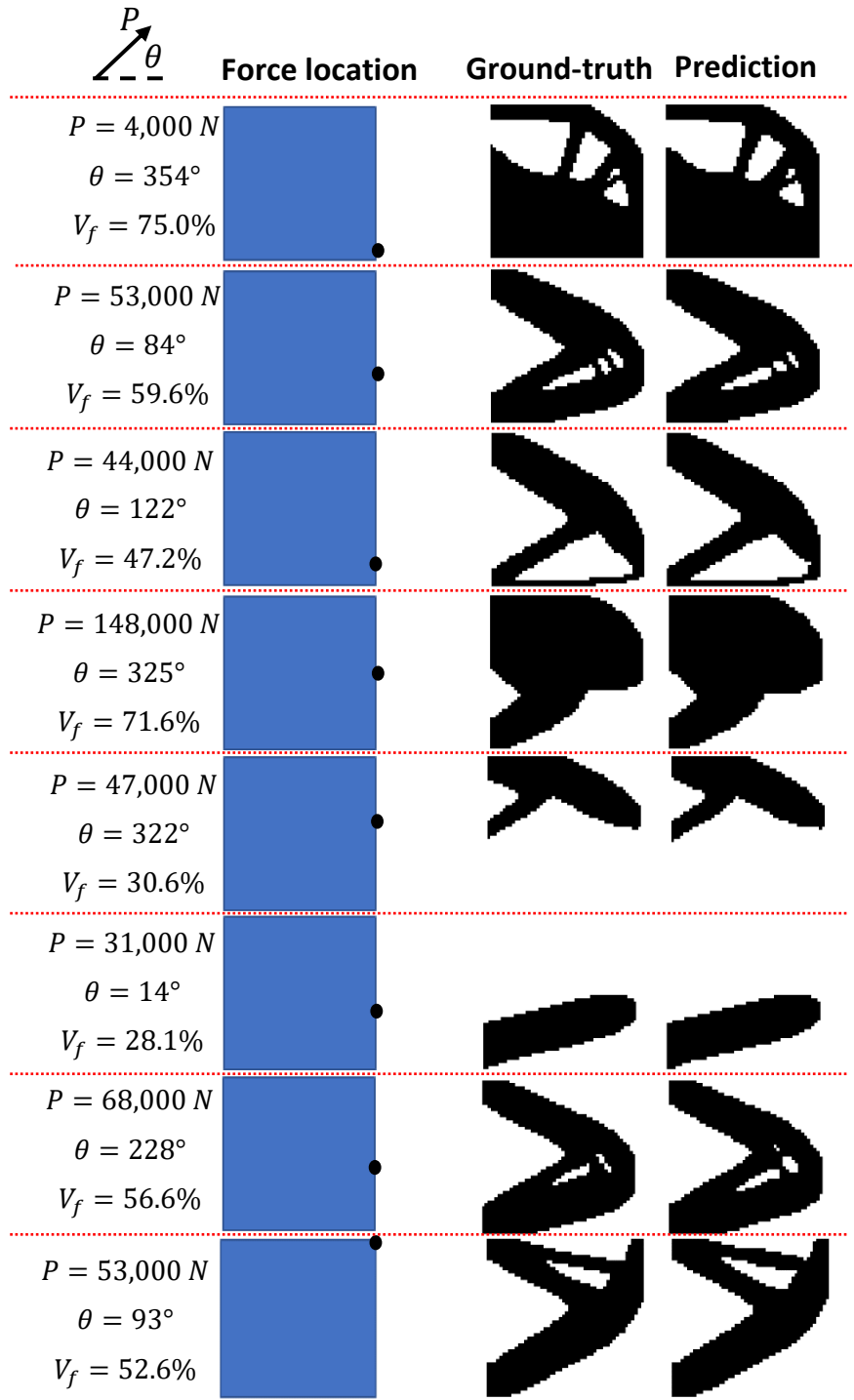

Figure 11: Comparison between optimized designs for the case of nonlinear elasticity with finite deformation. The design space has a dimension of $1m \times 1m$.

For the sake of qualitative evaluation of the model, we pick random ground-truth designs and their corresponding predicted ones and compare them. Figure 10 shows some examples of ground-truth and predicted designs. The results obtained from the developed CNN model are almost identical to the ground-truth results. Also, in the work of Yu et al. [71], the model used provided some structural disconnections in some cases, while such disconnections did not appear in the ground-truth data. This implies that there is a kind of discrepancy in the developed model. In the present paper, the same number of elements ($32 \times 32$) has been considered, and the issue of structural disconnections is not encountered, although fewer data points (15,000 data points compared to 100,000 data points) are used to train our model. Similar structural disconnections are also observed in the work of Zhang et al. [74]. Although the architecture of the ResUnet is more complex than those of conventional CNN models used for topology optimization problems, this complexity results in more accurate model trained on a relatively small dataset. One factor that leads to such a robust performance of the developed model is the combination of low-level information and high-level information. Figure 5 shows this information transfer from the encoder to the decoder. In addition to the architecture of the ResUnet, the random generation of data might also have led to the robust performance of the model.

Next, we show the results obtained from the model developed for the nonlinear case, neo-Hookean material with finite deformation. Figures 8b and 9b show the convergence history of the loss function and $DSC$, respectively. It can be implied from these figures that no major overfitting is occurring. One method to avoid overfitting is early stopping [100-102], a form of regularization. Here, we early stop the training process at 150 epochs. After the training process is completed, the ground-truth and prediction images in the testing dataset are compared; the mean $DSC$ for the testing dataset is 0.964. Figure 11 portrays a few examples of ground-truth and predicted designs. The results obtained from the developed CNN model almost coincide with the ground-truth results.

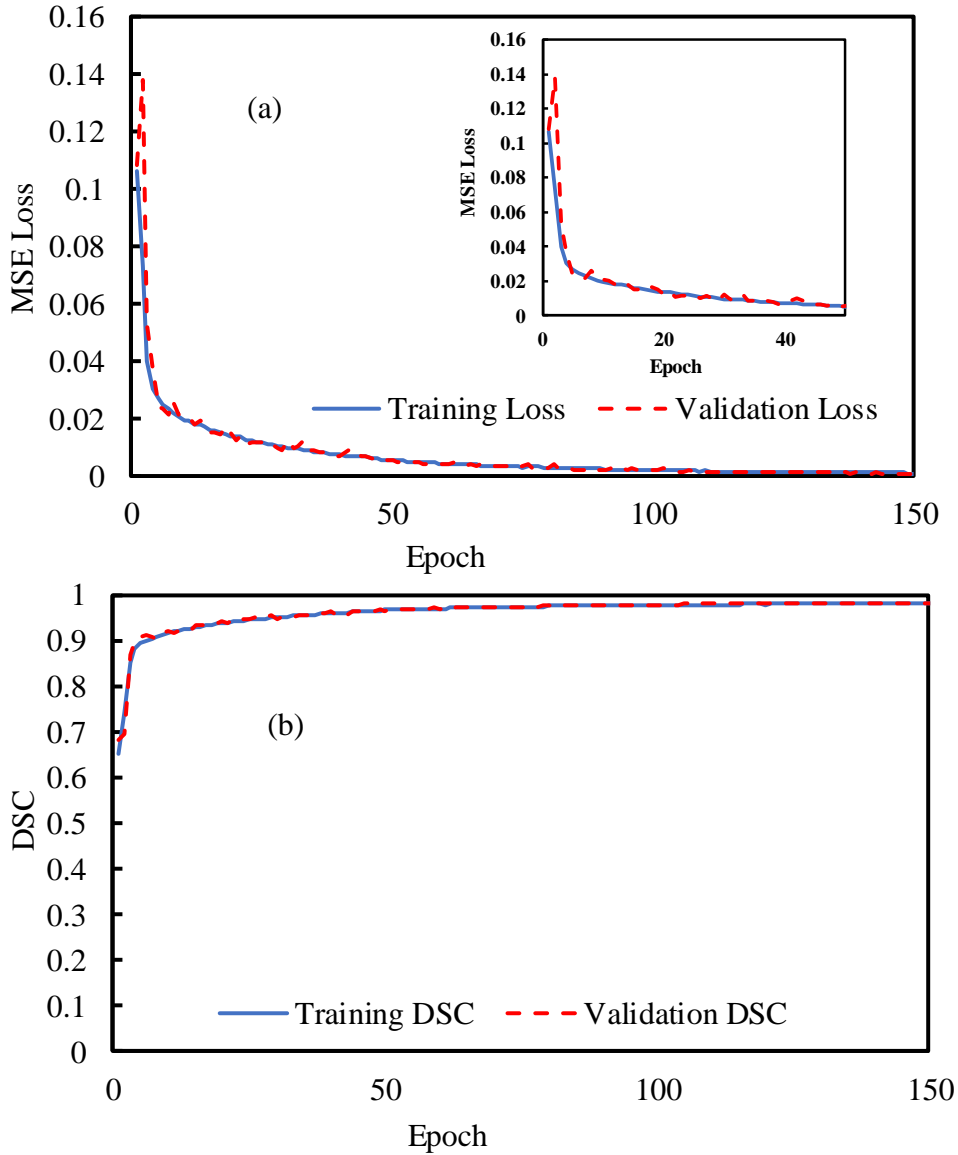

Figure 12: The convergence history of the CNN model developed for the case of linear elasticity under stress constraint: (a) loss history and (b) $DSC$ history.

Then, we present the case of linear elasticity under stress constraint. Figure 12 depicts the convergence history of the loss function and $DSC$ of the developed CNN model. It is inferred from the figure that the CNN model is robust. After the completion of the training process, the prediction and ground-truth images in the testing dataset are compared; the mean $DSC$ for the testing dataset is 0.984. Figure 13 shows a comparison between the ground-truth and predicted designs for a few examples. The results obtained from the developed CNN model almost coincide with the ground-truth results.

Also, the proposed framework can be generalized to arbitrary design spaces by adding an extra input channel defining the geometry of design spaces, prescribed displacements at different locations, and/or multiple loads (or even uniform load) leading to a multipurpose machine learning model for topology optimization. The proposed framework can be applied to other material nonlinearities such as plasticity and viscoplasticity with or without geometric nonlinearities. Additionally, one can use generative adversarial networks to refine the resolution [71]. The ability to generalize to scenarios discussed above requires data accounting for the different cases. Otherwise, such data-driven topology optimization models would lack the ability to generalize for scenarios that are not accounted for during the training process. There were a few attempts to generalize such models by using different input channels, as discussed in the work of Zhang et al. [74].

After the training of the machine learning model is complete on HPC, the trained learnable parameters (weights and biases) can be transferred to any low-end computing platform such as a laptop, and the optimized solutions are found there instantly without any iterations for any variation of input parameters. The proposed data-driven method can almost instantly provide preliminary optimized designs, thus quickly guiding to initial designs for subsequent conventional topological optimization and significantly increasing convergence rate and computational and overall efficiencies of the design process. As the higher-end hardware becomes more available and affordable while the machine learning methods further mature and their confluence becomes more widely accepted by the computational mechanics communities, we believe that data-driven models will pave the way for remarkably efficient design and modeling with topology optimization and other computationally intensive numerical methods.

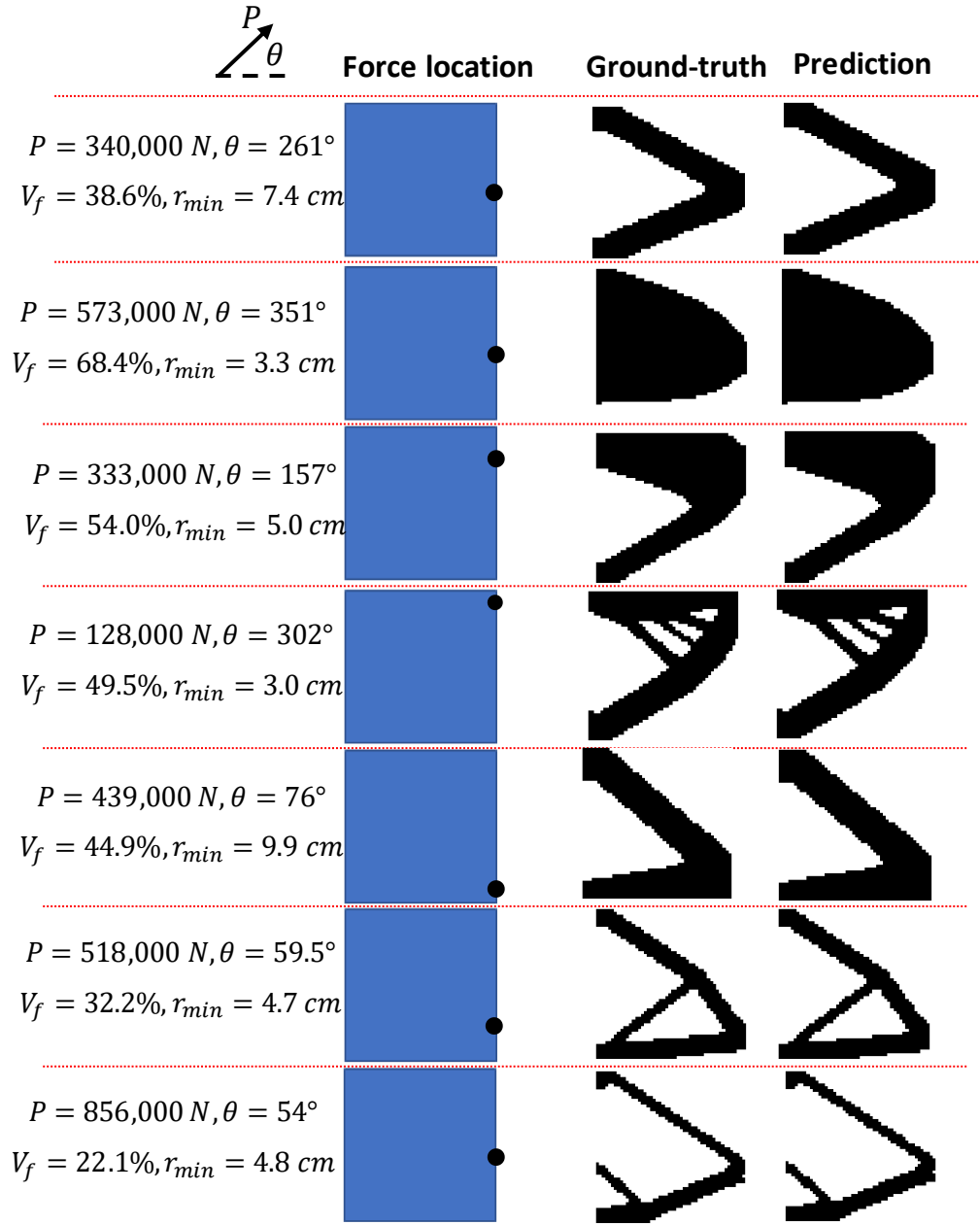

Figure 13: Comparison between optimized designs for the case of linear elasticity under stress constraint. The design space has a dimension of $1m \times 1m$.

## 6. Conclusions and future work

In this paper, we develop three CNN models to predict the optimized designs in the case of a) linear elasticity with small deformation (without nonlinear constraints), b) nonlinear hyperelasticity (neo-Hookean material) with geometric nonlinearity, and c) linear elasticity with stress constraint, a nonlinear constraint. The developed machine learning models are robust, and they are in an excellent agreement with the results obtained from the mathematically rigorous nonlinear topology optimization frameworks, which require an expensive computational cost. We show that it is possible to generate, machine train, test, and predict data on HPC, and then instantly inference good quality nonlinear topology optimization results on a low-end computing platform such as laptops, which can quickly guide to preliminary designs. In future work, we will work on strengthening the generalization ability of the developed data-driven topology optimization model, including a 3D generalization with many 2D slices representing a 3D geometry, where 3D convolutional layers can be used.

## Acknowledgments

The authors would like to thank the National Center for Supercomputing Applications (NCSA) Industry Program for software and hardware support.

## Data Availability

The data that support the findings of this study are available from the corresponding author upon reasonable request.